\documentclass[prx,a4paper,twocolumn,noshowpacs,citeautoscript,nofootinbib]{revtex4-1}

\usepackage{amssymb}
\usepackage{graphicx}
\usepackage{amsmath}
\usepackage[english]{babel}

\usepackage{color}

\begin{document}

\title{Propagation of self-localised Q-ball solitons in the $^3$He universe}

\author{S. Autti$^{1}$}
\email{samuli.autti@aalto.fi}
\author{P.J. Heikkinen$^{1}$}
\altaffiliation{Present address: Department of Physics, Royal Holloway, University of London, Egham, Surrey TW20 0EX, UK}
\author{G.E. Volovik$^{1,2}$}
\author{V.V. Zavjalov$^{1}$}
\author{V.B. Eltsov$^{1}$}

\affiliation{$^{1}$Low Temperature Laboratory, Department of Applied Physics, Aalto University, POB 15100, FI-00076 AALTO, Finland}
\affiliation{$^{2}$Landau Institute for Theoretical Physics, 142432, Chernogolovka, Russia.}

\begin{abstract}
 In relativistic quantum field theories, compact objects of interacting bosons can become stable owing to conservation of an additive quantum number $Q$. Discovering such $Q$-balls propagating in the Universe would confirm supersymmetric extensions of the standard model and may shed light on the mysteries of dark matter, but no unambiguous experimental evidence exists. We report observation of a propagating long-lived $Q$-ball in superfluid $^3$He, where the role of $Q$-ball is played by a Bose-Einstein condensate of magnon quasiparticles. We achieve accurate representation of the $Q$-ball Hamiltonian using the influence of the number of magnons, corresponding to the charge $Q$, on the orbital structure of the superfluid $^3$He order parameter. This realisation supports multiple coexisting $Q$-balls which in future allows studies of $Q$-ball dynamics, interactions, and collisions.
\end{abstract}
\maketitle

\section{Introduction}

All self-bound macroscopic objects encountered in everyday life or observed experimentally are made from fermionic matter, while bosons mediate interactions between fermionic particles. Compact objects made purely from interacting bosons may, however, be stabilised in relativistic quantum field theory by conservation of an additive quantum number $Q$. \cite{coleman_qball,Cohen1986301,friedberg_qball}. Observing such $Q$-balls travelling in the Universe would have striking consequences: their discovery would support supersymmetric extensions of the Standard Model \cite{Kusenko199846,Enq}, $Q$-balls could have participated in baryogenesis \cite{Kasuya2} and formation of boson stars \cite{BosonStarPRD.77.044036}, and the dark matter \cite{Kasuya,Kus2,Enq,Kus} and supermassive compact objects in galaxy centres \cite{Troitsky2016} may consist of $Q$-balls. Nevertheless, unambiguous experimental evidence on $Q$-balls has so far not been found in cosmology or in high-energy physics. Here we provide a laboratory realisation of a long-lived $Q$-ball and observe its propagation in three dimensions. The $Q$-ball Hamiltonian is simulated using a Bose-Einstein condensate of magnon quasiparticles in superfluid $^3$He.  

Bose-Einstein condensation (BEC) of quasiparticles --- such as magnons \cite{magnon_BEC_review,Demokritov2006430}, exciton-polaritons \cite{Kasprzak2006409}, or even photons \cite{Klaers2010545} ---  keeps extending the limits of known macroscopic coherent phenomena \cite{magnonBEC_RT}. Quasiparticle condensates provide a perspective platform for experimental studies of elusive systems and exotic theoretical models, based on the tradition of quantum simulations in atomic BECs \cite{RevModPhys.80.885,RevModPhys.86.153}. One of the most versatile environments is provided by the superfluid phases of $^3$He, where a number of concepts from high energy physics and cosmology have already been successfully tested \cite{VolovikBook,Zurek1985,KZ_Nature,BraneAnnihilation,HiggsNComm}.
We use magnon BEC in the B phase of superfluid $^3$He to study formation and propagation of $Q$-balls. Magnons in $^3$He-B are quanta of transverse spin waves, accompanied by precessing magnetisation of $^3$He nuclei. Magnon condensation is manifested in spontaneous phase coherence of the precession \cite{magnon_BEC_review,1984_hpd_exp,1984_hpd_theor,1992_ppd,2000_ppd,2012_ppd}. 

In experiments the magnon BEC is trapped in a potential created by external magnetic field and by the spin-orbit interaction owing to
the spatial distribution of the orbital anisotropy axis $\hat{\bf l}({\bf r})$ of the superfluid $^3$He-B order parameter. At temperatures below $\approx 3\cdot 10^{-4}\,$K the lifetime of such condensates rapidly increases and reaches minutes \cite{2012_ppd,magnon_relax}. Unlike the common case of trapped atomic condensates, the magnon BEC modifies the underlying $\hat{\bf l}({\bf r})$ profile and, hence, the confining potential \cite{magnon_trap_mod}, providing a possible laboratory realisation of a $Q$-ball \cite{volovik_bunkov_qball}. Other condensed-matter analogues of $Q$-balls have been proposed earlier \cite{Hong1988,EnqvistLaine}, and properties of bright solitons in 1D atomic BECs \cite{1D_lithium_Qball} and Pekar polarons in ionic crystals \cite{DevreesePolaron} have some similarities with $Q$-balls. We show that the magnon BEC in $^3$He-B allows for the most accurate representation of $Q$-ball properties in three dimensions. We report experiments on a spontaneously moving $Q$-ball, initially formed on the periphery of the sample container in a self-created broken-symmetry trap. The peripheral trap gradually transforms into the trap in the centre of the sample, as favoured by the $^3$He order parameter, conforming the movement of the magnon BEC. This propagation unambiguously demonstrates the non-trivial soliton nature of a true, long lived $Q$-ball. 

 \begin{figure*}[!htb]
\centerline{\includegraphics[width=0.8\linewidth]{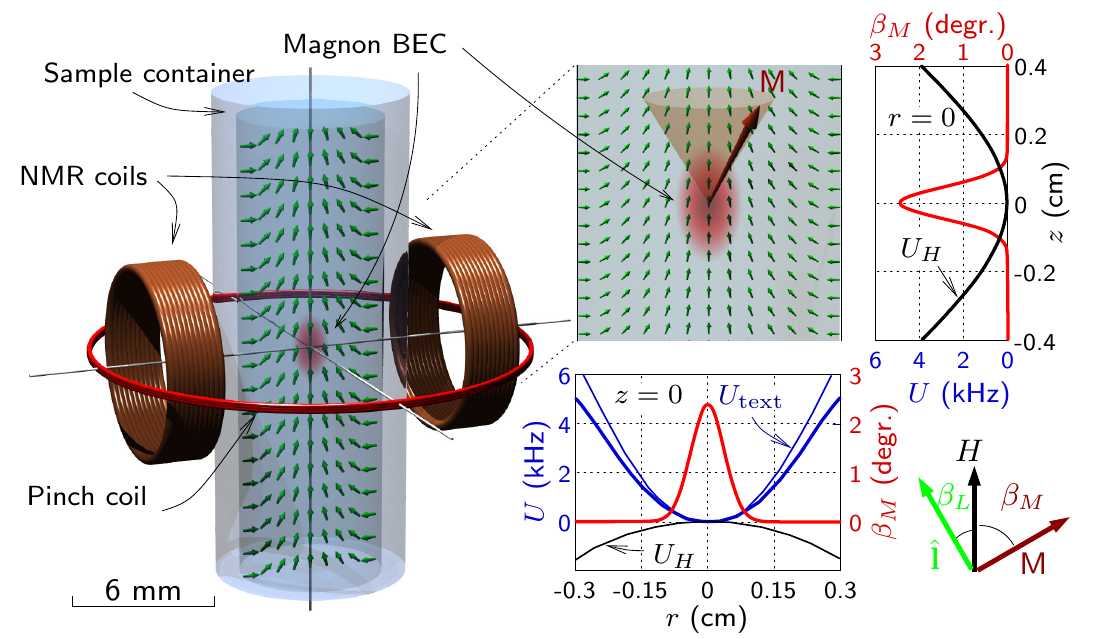}}
\caption{\label{image:01cell} (Color) Experimental setup: Top part of the sample container, magnon condensate with precessing magnetisation $\mathbf M$ (\emph{red blob}) in a potential trap (\emph{blue lines}), and corresponding wave functions for a small number of magnons (\emph{red lines}). In the radial direction potential minimum is formed by a combination of magnetic and textural energies, $U_H$ and $U_{\rm text}$. In the axial direction the minimum is formed by the magnetic energy alone. \emph{Green arrows} illustrate the spatial distribution of vector $\hat{\mathbf l}$, which is uniform in the $\hat{\mathbf z}$ direction in the absence of magnons. The pinch coil position defines $z=0$, which corresponds also to the common axis of the NMR pick-up coils. The potentials and wave functions are calculated for $T=0.15~T_\mathrm{c}$, $p=0.5$\,bar, and pinch coil current 3\,A.}
\end{figure*}

\section{Magnon BEC as $Q$-ball}

The essential component of a $Q$-ball is the relativistic complex field $\Phi$ of self-localised charge \cite{coleman_qball}. In the class of soliton solutions \cite{friedberg_qball}, which our experiment realises, the $\Phi$ field interacts with the neutral scalar field $\zeta$, which provides a confining potential. For a magnon $Q$-ball in superfluid $^3$He the field $\Phi$ is the transverse component of the coherently precessing spin, $\Phi \propto S_x+iS_y$. Quasi-conserved number of magnons, $N_{\rm M}$, becomes the $Q$-charge. The $\Phi$ field in a $Q$-ball obeys a relativistic Klein-Gordon equation \cite{coleman_qball,friedberg_qball}. In Appendix A we derive this equation for our magnon representation of $\Phi$ starting from the Leggett equations of spin dynamics in $^3$He-B (Eq. (\ref{dot_theta})). We show that in the long-wavelength limit realised in the experiments it transforms to a Schr\"odinger equation
 \begin{equation}
-i \hbar\frac{\partial \psi}{\partial t}= - \frac{\hbar^2}{2m}\Delta \psi + U({\bf r})\psi\,,
\label{SchrodingerEq}
\end{equation}
where $m$ is the magnon mass and $|\psi|^2 \propto |\Phi|^2$. The trapping potential $U({\bf r})$ is formed by the magnetic field $|\mathbf{H}({\bf r})|= \frac{2\pi}{|\gamma|} \nu_L({\bf r}) $ and the neutral field $\zeta$:
 \begin{equation}
U({\bf r})=U_H + U_{\rm text} \equiv h\nu_L({\bf r})+ \frac{1}{2\pi\nu_L}\zeta^2({\bf r}) \,.
\label{U}
\end{equation}
Here $\nu_L$ is the  Larmor frequency and $\gamma$ the gyromagnetic ratio of $^3$He. It turns out that the neutral field $\zeta({\bf r})$ can be expressed in terms of $\beta_L$, the deflection angle of $\hat{\bf l}$ measured from $\mathbf{H} \parallel  \hat{\mathbf{z}}$ (see  Eq. (\ref{TexturalPotential2})):
\begin{equation}
\zeta^2({\bf r}) \propto
\sin^2( \beta_L({\bf r}) / 2)
\,.
 \label{OrbitalField}
\end{equation}
The condensate wave function $\psi$ is normalised to the number of magnons $N_\mathrm{M}$, and can be expressed in terms of the tipping angle $\beta_{\rm M}$ of the precessing magnetisation (see  Eq. (\ref{psi_vs_precession_full})):
 \begin{equation}
\psi \propto \sin(\beta_{\rm M}(\mathbf{r})/2) e^{i\omega t} \,.
\label{psi_vs_precession}
\end{equation}
The frequency of precession $\omega=2\pi\nu$ plays the role of
chemical potential of the magnon BEC, and $t$ is time. For a detailed derivation of the above quantities, see Appendix A.

In the absence of magnons the spatial distribution of $\hat{\bf l}$ in our cylindrical container results from competing effects of the magnetic field and the container walls (see Fig~\ref{image:01cell}): The orientation changes smoothly from parallel to the field at the container axis to perpendicular to the wall at the periphery. Together with the magnetic potential, the profile of $\hat{\bf l}$ leads to a nearly harmonic 3D potential \cite{magnon_trap_mod}. We put origin of our coordinate system to the bottom of this trap and choose $U(r=0,z=0)=0$. Therefore the condensate energy is conveniently measured as the shift $\Delta \nu$ of the precession frequency of the magnetisation from the Larmor frequency at the origin: $\nu=\nu_L(r=0,z=0)+\Delta \nu$.  All magnon states in this harmonic trap, including the ground state, have the frequency shift $\Delta\nu>0$. Relevant parts of the sample container, an example of the trapping potential, and the corresponding condensate wave function are shown in Fig.~\ref{image:01cell}.

The $Q$-ball Hamiltonian in general contains a repulsive interaction between the charged and neutral fields. Here it arises from the spin-orbit interaction, which increases free energy by $F_{\rm so}=|\Phi({\bf r})|^2 \zeta^2({\bf r})$. As the number of magnons increases, $\hat{\bf l}$ within the condensate reorients along $\mathbf{\hat{z}}$, reducing $U_{\rm text}({\bf r})$ and the energy eigenstate in the trap. Experimentally this is observed as decrease of the condensate precession frequency $\nu$ with increasing signal amplitude. At large $N_{\rm M}$ the effect becomes so strong that $U_{\rm text}({\bf r})$ forms a box \cite{magnon_trap_mod} with a flat bottom and steep walls. This box is a bosonic analogue of a hadron in the MIT bag model \cite{rota2012}, as elaborated in Appendix B, and an essential prerequisite for formation of a $Q$-ball.

\section{Observing $Q$-balls in experiments}

In our experiments the superfluid $^3$He sample, contained in a long cylindrical quartz tube (diameter~5.8~mm, length~150~mm), is cooled down using a nuclear demagnetisation refrigerator to (0.13--0.20)~$T_c$. The superfluid transition temperature $T_\mathrm{c}=1$\,mK at pressure $p = 0.5$\,bar used here (if not specified otherwise). Temperature is measured using a quartz tuning fork sensitive to the thermal quasiparticle density in the sample \cite{2007_forks, 2008_forks}. The fork is located near the bottom of the container above the sintered connection to the nuclear demagnetisation cooling stage. The applied magnetic field is 25.4~mT, and the corresponding nuclear magnetic resonance (NMR) frequency $\nu_L=\omega_L/2\pi= 826$~kHz. In addition to the homogeneous axial field used for NMR, we use a pinch coil to create a field minimum along the sample container axis centred at $z=0$. The pinch coild produces also in a small field maximum in the radial direction. The experimental setup and the magnetic field profile calibration are described in more detail in Refs.~\cite{Zavjalov2015, magnon_relax}. 

To monitor the formation and propagation of $Q$-balls we use NMR techniques. They have proved powerful in probing various phenomena in $^3$He close to zero temperature \cite{magnon_relax, texture_vortex}. Magnons are pumped to the system with a radio-frequency pulse at a frequency above the ground state frequency. The pumped magnons then quickly condense to the ground state forming the BEC \cite{magnon_BEC_review}. The coherently precessing magnetisation of the condensate induces signal in the NMR pick-up coils with amplitude $A\propto\int \sin \beta_M\,dV$. Frequency and amplitude of the recorded signal are extracted as a function of time by tracing the peak in a windowed Fourier transformation of the signal. For a fixed geometry of the condensate $A \propto N_{\rm M}^{1/2}$, but the proportionality coefficient depends strongly on the spatial distribution of the BEC wave function. This allows us to track location of the $Q$-ball in the measurements. 

\begin{figure}[t!]
\centerline{\includegraphics[width=\linewidth]{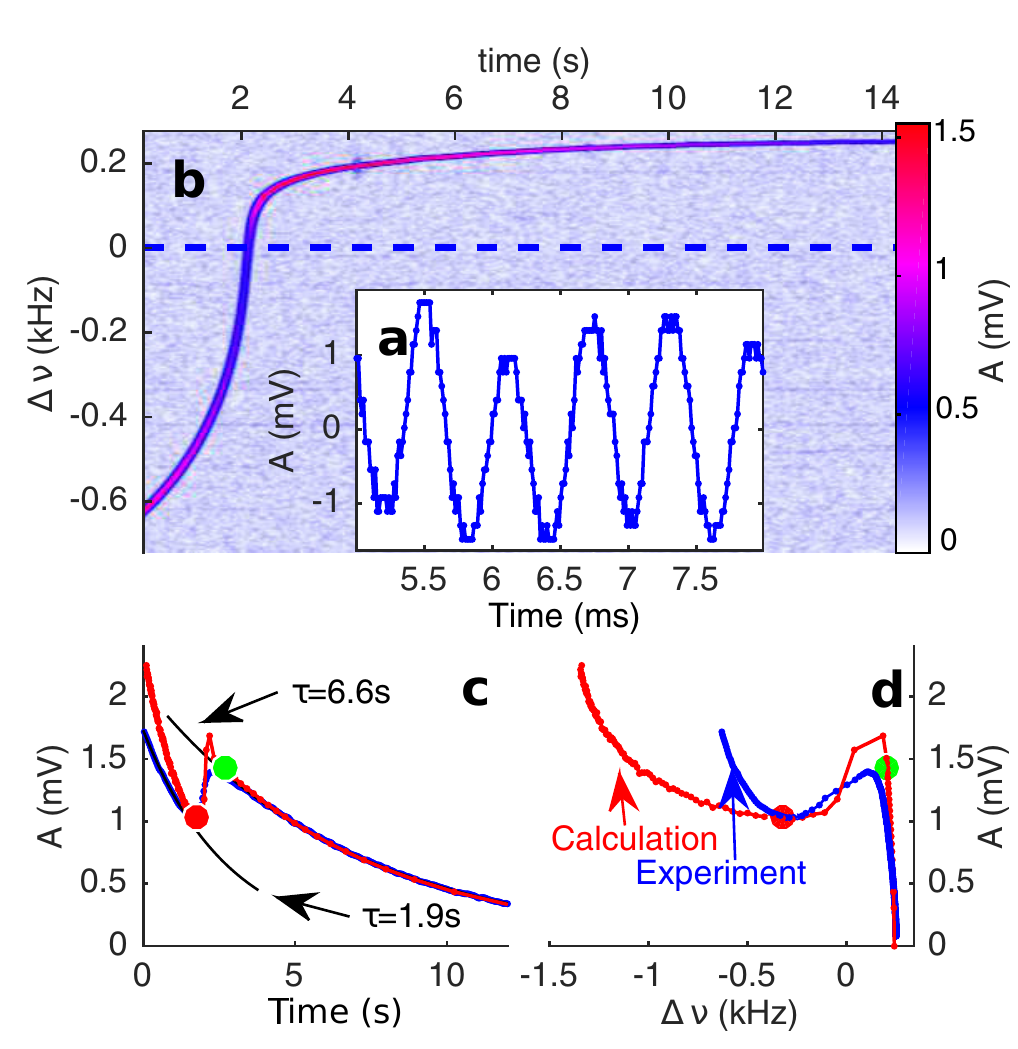}}
\caption{\label{image:02pulse} (Color) Frequency and amplitude of a magnon $Q$-ball during its decay: The $Q$-ball is created by an RF pulse at $t=0$ ($T=0.15\,T_\mathrm{c}$, $T_\mathrm{c}\approx 1$mK). ({\textbf a}) The signal recorded from the NMR pick-up coils. ({\textbf b}) Windowed Fourier transform of the signal showing the magnon BEC as a sharp peak whose frequency shift $\Delta \nu$ and amplitude $A$ change in time. ({\textbf c}\&{\textbf d}) Measured (\emph{blue lines}) and calculated (\emph{red lines}) dependencies $A(t)$ and $A(\Delta \nu)$. \emph{Black lines} in ({\textbf c}) are exponential fits to the measured decay. The initial decay time in the simulation is $\tau = 1.7\,\rm{s}$. \emph{Red} and \emph{green circle} mark the peripheral and central $Q$-balls illustrated in Fig. \ref{image:03interpretation}.}
\end{figure}

\begin{figure}[htb!]
 \centerline{\includegraphics[width=1\linewidth]{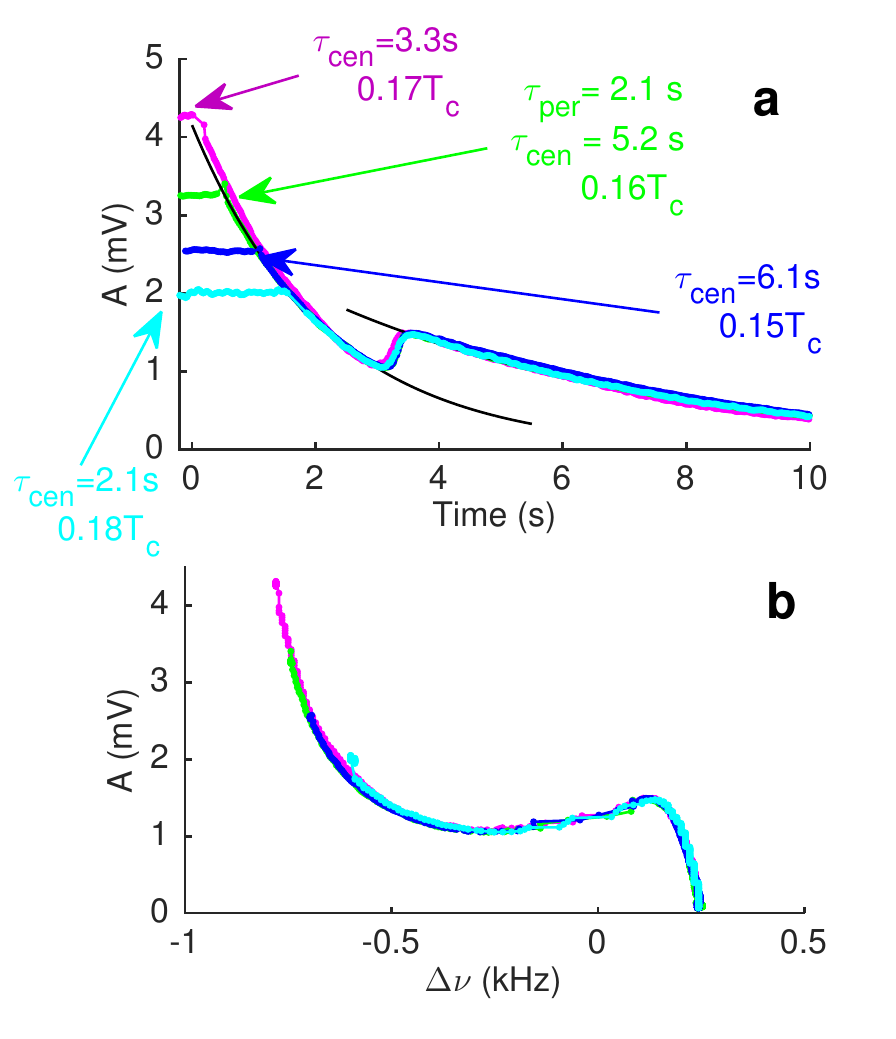}}
   \caption{\label{Fig:temps_and_amps} (Color) ({\bf a}) $Q$-ball decays and ({\bf b}) corresponding frequency-amplitude dependencies at different temperatures and with different initial amplitudes at $P=0$~bar. The measured decay signals are scaled in time according to $\tau_{\rm cen}$ to the units of the \emph{green line}, revealing that the decay paths are identical in both $A(t)$ and $A(\Delta \nu)$. The signals begin with a non-decaying part of constant amplitude, where the condensate is supported by RF pumping ($t<0$ in Fig.~\ref{image:02pulse}). In this plot $t=0$ corresponds to turning off pumping for the \emph{magenta line}, and other lines have been shifted in time in order to allow comparing the decay processes. $\tau_{\rm cen}$ stands for the time constant of exponential decay of the central $Q$-ball (tail of the signal), and  $\tau_{\rm per}$ for that of the peripheral one (beginning part of the signal).  \emph{Black lines} are exponential fits to the \emph{green line}. }
\end{figure}

\section{Propagating $Q$-ball}

After the exciting pulse is turned off at time $t=0$, $N_{\rm M}(t)$ decays slowly due to non-hydrodynamic spin diffusion and radiation damping \cite{magnon_relax}, the former being the dominant contribution. If at $t=0$ the number of magnons $N_{\rm M}(0)$ is relatively small, the signal amplitude decays exponentially, $A(t) \propto\exp(-t/\tau)$, and the change in the frequency shift $\Delta\nu(t)$ during the decay is small \cite{magnon_trap_mod,magnon_relax}. With high $N_{\rm M}(0) > N_{\rm M}^{\rm c} \sim 10^{12}$, we observe reproducible decay signals with non-monotonous $A(t)$ and $\Delta\nu(0) < 0$, that is, $\nu$ is below the minimum of the original trapping potential, Fig.~\ref{image:02pulse}. The relaxation process is well defined: The relaxation follows a sequence of states which is independent of the relaxation rate, controlled by temperature, as demonstrated in Fig.~\ref{Fig:temps_and_amps}a. Decays started from different $N_{\rm M}(t)$ are identical after the common signal amplitude is reached, see Fig.~\ref{Fig:temps_and_amps}b.

\begin{figure}[tb!]
\centerline{\includegraphics[width=\linewidth]{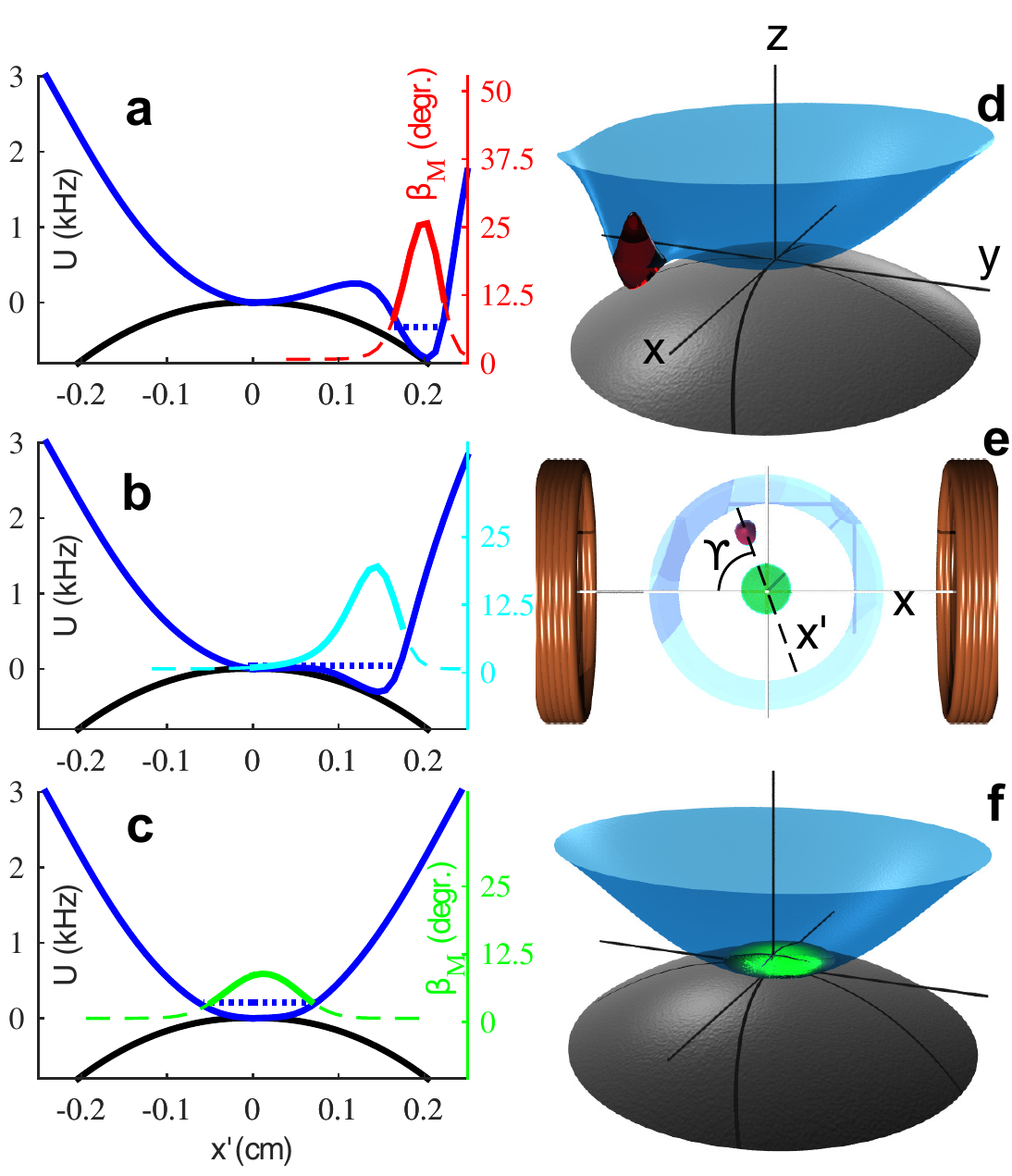}}
\caption{\label{image:03interpretation} (Color) Propagating magnon $Q$-ball: ({\textbf a \& \textbf c}) The peripheral (\emph{red line}) and central $Q$-balls (\emph{green line}), as marked in Fig~\ref{image:02pulse}, are plotted in terms of $\beta_M$ at $z=0$ along the direction of $Q$-ball movement, labelled $x'$. \emph{Solid line} is used where the condensate frequency (level indicated by \emph{dotted blue line}) is above the total potential $U$ (\emph{solid blue line}), and \emph{dash line} where it is below. The magnetic potential $U_H$ is drawn with \emph{solid black line}. ({\textbf b}) While propagating from the periphery to the axis (\emph{light blue line}), the $Q$-ball frequency crosses $\Delta \nu =0$. ({\textbf d} \& { \textbf f}) The parts of the peripheral and central $Q$-ball wave functions (\emph{red} and \emph{green surfaces}) that correspond to $Q$-ball frequency being above the potential (\emph{blue surface}) are plotted in the $z=0$ plane to illustrate the broken azimuthal symmetry of 
the peripheral state. ({\textbf e}) Top view of the sample container, the NMR coils, and the two $Q$-ball states (as plotted in {\textbf d} \&{ \textbf f}) reveals the time evolution of the $Q$-ball size. The peripheral $Q$-ball is plotted travelling to one of the four degenerate directions w.r.t. the NMR coils.}
\end{figure}

We explain these observations via formation of a peripheral magnon $Q$-ball in a spontaneous broken-symmetry trap: Our self-consistent numerical simulation (details below) shows that with a sufficiently large $Q\equiv N_{\rm M}$ the textural potential $U_{\rm text}$ is suppressed due to the above-mentioned box effect. The radial maximum in the magnetic potential $U_H$ allows the $Q$-ball to self-localise in the periphery. Remarkably, the axial symmetry of the confinement is spontaneously broken as well and the $Q$-ball becomes also azimuthally localised. This phenomenon is unlike conventional spontaneous symmetry breaking where the potential remains axisymmetric (Appendix C). At the periphery of the sample, the $Q$-ball's energy is below the original trap minimum and thus $\Delta\nu < 0$. In the simulations the $Q$-ball moves further from the axis (that is, closer to the sample container wall) than in the experiments, and hence its frequency is lower. This is probably due to insufficient rigidity in the model's orbital texture, which keeps the $Q$-ball from eventually colliding with the container wall. The simulation is compared with the experiment in Fig.~\ref{image:02pulse} and interpreted in Fig.~\ref{image:03interpretation}.

The soliton nature of the propagating $Q$-ball is manifested during the decay: The $Q$-ball decays while staying at the periphery until it reaches the critical charge $Q_{\rm{c}}=N_{\rm M}^{\rm c}$, after which it quickly propagates to the centre and simultaneously changes shape. This is an experimental realisation of the threshold $Q_{\rm{c}}$ discussed in Ref. \cite{friedberg_qball}. Thereafter the exponential decay continues roughly at three times slower rate. The non-monotonic evolution of the signal amplitude is a signature of the propagation: The peripheral $Q$-ball is strongly localised, that is, compressed in both azimuthal and axial directions due to pressure of the surrounding texture. The central $Q$-ball spreads wider (see Fig. \ref{image:03interpretation}e) and therefore produces a larger signal for a given number of magnons. On top of the relatively slow decay of $N_{\rm M}(t)$, the fast propagation is therefore seen as a sudden increase in the signal amplitude. The change in the wave function also explains the different relaxation rates of the peripheral and central $Q$-balls: The relaxation is mainly due to spin transfer over the thermal quasiparticles in $^3$He-B, which increases with gradients of the wave function \cite{magnon_relax}. Those are larger for the compressed peripheral state. 

In the simulations we treat the decay of the $Q$-ball as a sequence of quasi-equilibrium states. This assumption is justified by the fact that in the experiments the observed sequence of states along the $Q$-ball decay is relaxation-rate independent (Fig.~\ref{Fig:temps_and_amps}). The limitations of this approach are revealed in the modest overshoot in simulated signal amplitude when the $Q$-ball moves to the centre (Fig.~\ref{image:02pulse}~b,c). We solve for the charged field (Eq.~(\ref{SchrodingerEq})) and the neutral $\hat{\bf l}$-field for each $N_\mathrm{M}$, varying $N_\mathrm{M}$ in steps. Self consistency between $\psi$ and $\hat{\bf l}$ is reached with a fixed point iteration. Close to $Q_{\rm{c}}$ the fixed point iteration becomes sensitive to the initial condition. We start the simulation from $N_\mathrm{M} \gg Q_{\rm{c}}$, and use the solution at the previous step as initial condition for the next step. The $\hat{\bf l}$-profile is calculated in 3D by minimisation of appropriate free energy \cite{thuneberg_texture, kopu_texture} including interaction with the magnon condensate in Eq.~(\ref{InteractionIntroduction}). Solving Eq.~(\ref{SchrodingerEq}) when $N_M=0$ is 
described in Ref.~\cite{magnon_relax}. Time evolution of the $Q$-ball in simulations is calculated by solving Eq.~(7) in Ref.~\cite{magnon_relax} for the relaxation rate of the Zeeman energy. The signal amplitude in simulations is scaled to fit the very tail of the decay of the experimental signal, where the decay is well understood \cite{magnon_relax}.

The expected NMR signal from the precessing magnetisation in the simulation is calculated using the known geometry of the coil system. The direction that the peripheral $Q$-ball is moving to, that is, the angle $\Upsilon$ between axes $x$ and $x'$ in Fig. \ref{image:03interpretation} is fitted, yielding $\Upsilon=67^\circ$. This fitted value of $\Upsilon$ corresponds to four possible directions of $Q$-ball's movement due to the symmetry of the coil system. Closer to the coils their sensitivity is higher and, hence, the peripheral $Q$-ball would produce roughly twice larger signal than observed, should it travel towards one of the two coils ($\Upsilon=0^\circ$), and about 50\% smaller signal if it travelled along the direction perpendicular to the common axis of the NMR coils ($\Upsilon=90^{\circ}$). The signal produced by the central condensate does not depend on $\Upsilon$. To control this symmetry breaking in the simulation, we introduce a small symmetry violating perturbation in the simulated potential to lift the 
degeneracy without influencing the structure of the $Q$-ball.

\section{Coexistence of two $Q$-balls}

\begin{figure}[h!]
\centerline{\includegraphics[width=1.05\linewidth]{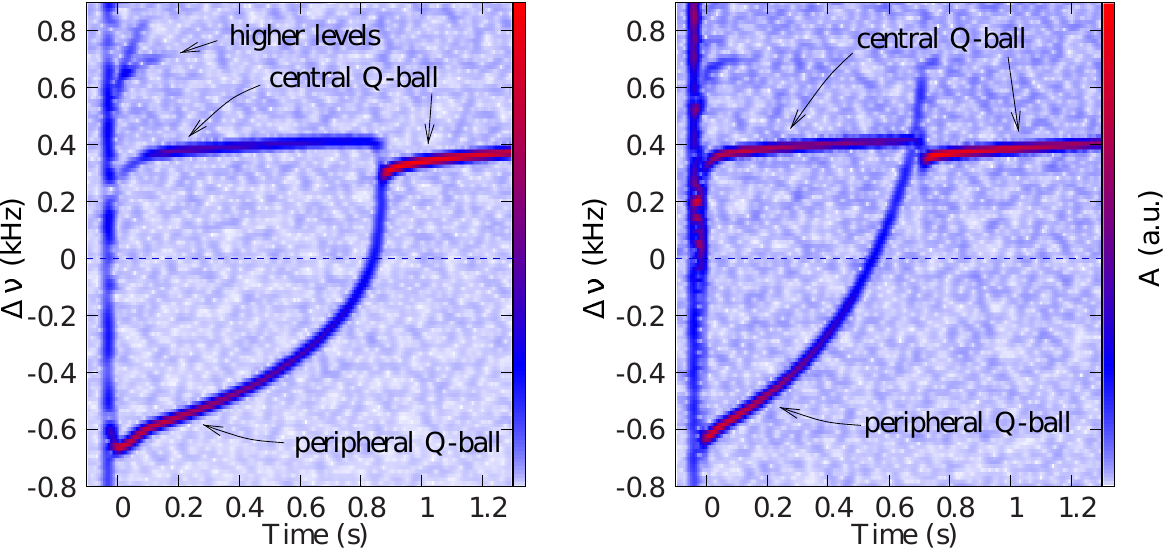}}
\caption{\label{image:04pulse2st} (Color) Coexistence of central and peripheral $Q$-balls: a true $Q$-ball on the periphery of the sample container in the broken-symmetry trap, and a central $Q$-ball on the sample container axis (1~rad/s rotation, $T$=0.13~$T_c$). During the decay both eigenenergies (frequencies) increase and when they become close enough, the condensates merge. This process of merging of two $Q$-balls into a single one is demonstrated in two different experiments conducted under similar conditions ({\it left} and {\it right}). Figure on the left shows straightforward merger, whereas on the right the peripheral $Q$-ball goes through a meta-stable state: just before the merger its energy is higher than that of the central $Q$-ball. The meta-stability of the $Q$-balls obtained by interaction of charged and neutral fields is discussed in Ref.~\cite{friedberg_qball}. Close to $t=0$ some exited levels in the central trap are visible at higher frequencies.}
\end{figure}

The spatial distribution and rigidity of neutral field $\zeta(\mathbf{r})$ can be controlled by adding an array of quantized vortices by rotating the sample \cite{texture_vortex}. Rotating at 1~rad/s, we are able to create two coexisting spatially separated $Q$-balls using a RF pulse with wide enough spectrum (see  Fig.~\ref{image:04pulse2st}). That is, in addition to the $Q$-ball on the periphery of the sample container, there is another $Q$-ball localised to the container axis. They are stable owing to increased rigidity of the neutral field separating them. Due to the magnetic field profile, the central $Q$-ball has higher energy than the one on the periphery. 

During relaxation the peripheral $Q$-ball moves towards the sample container axis, and when the energies of the two $Q$-balls are sufficiently close, they merge forming a single magnon $Q$-ball in the central trap. This process is not very regular  and depends on, e.g., the phases and the initial amplitudes of the magnon BECs. The coexistence of two magnon $Q$-balls will allow detailed studies of interactions between them, especially Josephson effect between two $Q$-balls in flexible traps \cite{BEC_Josephson}. In future this setup can also be used, e.g., for a quantum simulation of the Penrose-type ``gravitationally'' induced wave function collapse \cite{PhysRevLett.14.57}.

\section{Conclusions}

The concept of $Q$-balls originates from high energy physics and cosmology where it has been used for so-far speculative explanations of many important phenomena in the Universe, such as the dark matter. We present an experimental confirmation of the $Q$-ball concept in a three-dimensional quantum simulation using a Bose-Einstein condensate of magnon quasiparticles in superfluid $^3$He-B. This realisation relies on the long lifetime of the magnons and their interaction with the orbital degrees of freedom of the underlying superfluid system. Both these phenomena are also important from the point of view of BEC formation and spin superfluidity in general. The $Q$-ball provides a new manifestation of the spin superfluidity of magnon BEC in $^3$He-B, which complements the Josephson effect,  quantized vorticity, superfluid spin currents, and the propagating Goldstone mode observed earlier in such condensates \cite{magnon_BEC_review,Sonin_book}. 

In our experiment the $Q$-ball propagates over a macroscopic distance in the sample container, and the confining potential conforms that movement. Such movement manifests the soliton nature of a true $Q$-ball. We further demonstrate how this realisation provides the possibility of creating two co-existing $Q$-balls and observing how they interact and merge. Detailed study of dynamics and interaction between the $Q$-balls, such as the AC Josephson effect, remains an interesting task for the future. 

\vspace{-0.8cm}
\section*{Acknowledgements}
\vspace{-0.5cm}
We thank V.V. Dmitriev, P. Skyba, and J. Nissinen for stimulating discussions. This work has been supported by the Academy of Finland (project no. 284594). The work was carried out in the Low Temperature Laboratory, which is part of the OtaNano research infrastructure of Aalto University. P.J.H. acknowledges financial support from the V\"{a}is\"{a}l\"{a} Foundation of the Finnish Academy of Science and Letters and from the Finnish Cultural Foundation, and S.A. that from the Finnish Cultural Foundation. The work by G.E.V. and V.B.E.  has been supported by the European Research Council (ERC) under the European Union's Horizon 2020 research and innovation programme (Grant Agreement No. 694248).

\section*{Appendix A: Derivation of $Q$-ball representation by Magnon BEC}


\renewcommand{\thefigure}{A\arabic{figure}} 
\setcounter{figure}{0}
\renewcommand{\theequation}{A\arabic{equation}} 
\setcounter{equation}{0}

\subsection*{Magnons as relativistic particles}

In what follows we  derive the components of a magnon $Q$-ball in detail. For further discussion of these topics, please see Refs.~\cite{Volovik2008} and \cite{volovik_bunkov_qball}. 

Let us start from the the  spectrum of transverse spin wave modes in $^3$He-B in magnetic field,
\begin{equation}
 \omega_\pm (k)= 
\pm  \frac{\omega_L}{2}   +\sqrt{ \frac{\omega_L^2}{4} +k^2c^2} \,,
 \label{MagnonSpectrumextended}
\end{equation}
where $\omega_L=\gamma H$ and $c$ is the spin wave velocity, which here is assumed isotropic for simplicity.
The spectrum (\ref{MagnonSpectrumextended}) can be considered as a spectrum of a relativistic particle 
with spin $S_z=\pm \hbar$ in an effective magnetic field:
\begin{equation}
E (S_z,k)=\sqrt{E_0^2+k^2c^2} - \gamma\tilde H S_z~~,
 \label{MagnonSpectrumextendedM}
\end{equation}
Here the rest energy $E_0$ of a particle is defined through its mass $m$ as
  \begin{equation}
E_0=  m c^2 =\frac{\hbar\omega_L}{2} \,,
\label{IsotropicMass}
\end{equation}
and the effective magnetic field is
\begin{equation}
\gamma \tilde H=\frac{\omega_L}{2} \,.
 \label{EffectiveH}
\end{equation}
At small $k$ when $ck \ll \omega$, 
the spectrum transforms to that of the Galilean limit of a massive particle
\begin{align}
&E(S_z=-\hbar,k)= \hbar\omega_L   +\frac{\hbar^2 k^2}{2m} , \\ 
&E(S_z=+\hbar,k)= \frac{\hbar^2 k^2}{2m} \,.
\label{MagnonSpectrum3}
\end{align}
The mode with $S_z=-\hbar$ corresponds to optical magnons relevant to this work. They are called simply magnons throughout the text. Each magnon reduces the projection of spin on axis $z$ by $\hbar$. The other branch ($S_z=\hbar$) is known as acoustic magnons \cite{HiggsNComm}.

The effective magnetic field can be removed by transformation to the spin reference frame rotating  with angular velocity $\omega_L/2$. In this frame the spectrum of spin waves becomes
\begin{equation}
  \tilde E(k) =\sqrt{E_0^2+k^2c^2} \,.
 \label{MagnonSpectrumextendedRef}
\end{equation}
The relativistic spectrum of spin waves suggests that
magnons can be seen as quanta of a "relativistic" quantum field. Below we show that the field is a scalar field and magnons therefore play the role of the $\Phi$-field, which appears in high-energy physics as the core component of the $Q$-ball soliton. 
In what follows we put $\gamma=\hbar=1$. Where relevant, these quantities are however expressed explicitly. 

\subsection*{Deriving the Relativistic Spectrum}

Let us write the linearised Leggett equations for spin dynamics in terms of the small spin-rotation angle $\mbox{\boldmath$\theta$}$,  $|\mbox{\boldmath$\theta$}| \ll 1$, which is related to the deviation of spin density ${\bf S}$ from its equilibrium value $\chi {\bf H}$ \cite{VW}:
\begin{equation}
 \bf S -\chi {\bf H} =\chi\partial_t\mbox{\boldmath$\theta$}\,.
 \label{dot_theta}
\end{equation}
Here $\chi$ is the spin susceptibility.

The Lagrangian for the $\mbox{\boldmath$\theta$}$-field contains a linear term in time derivative, because the magnetic field violates time reversal symmetry:
\begin{equation}
L= \frac{\chi}{2} \left(-\left(\partial_t\mbox{\boldmath$\theta$}\right)^2 - 
\left(\mbox{\boldmath$\theta$} \times  \partial_t\mbox{\boldmath$\theta$} \right)\cdot{\bf H} +c^2\nabla_i \mbox{\boldmath$\theta$}\nabla_i \mbox{\boldmath$\theta$}
\right)  +F_{\rm so}( \mbox{\boldmath$\theta$})\,.
 \label{Lagrangian_theta}
\end{equation}
 The term $F_{\rm so}( \mbox{\boldmath$\theta$})$ is spin-orbit interaction. It originates from dipole-dipole interaction between spins of the particles forming a Cooper pair and violates spin rotation symmetry.
The  Lagrangian (\ref{Lagrangian_theta}) can be rewritten in the following way:
\begin{align}
L= &\frac{\chi}{2} \left(-\left(\partial_t\mbox{\boldmath$\theta$}
 +\frac{1}{2}\mbox{\boldmath$\omega$}_L\times \mbox{\boldmath$\theta$}\right)^2
+E_0^2 \mbox{\boldmath$\theta$}_\perp^2+c^2\nabla_i \mbox{\boldmath$\theta$}\nabla_i \mbox{\boldmath$\theta$}
\right) \nonumber \\ 
&+F_{\rm so}( \mbox{\boldmath$\theta$})\,.
 \label{Lagrangian_theta2}
\end{align}
If one ignores the spin-orbit coupling, this Lagrangian describes relativistic massive particles that interact with a $SU(2)$ gauge field, whose time component is ${\bf A}_0=\frac{1}{2}\mbox{\boldmath$\omega$}_L$ \cite{MineevVolovik1992,2008Zaanen}.

 Let us consider transverse NMR, where only the components 
$\mbox{\boldmath$\theta$}_\perp \perp {\bf H}$ are relevant (the static magnetic field is along the $z$-axis). 
 The gauge field is curvature free and can be eliminated, like above, by time dependent rotation in spin space, which corresponds to the transformation to the spin reference frame rotating with angular velocity $\omega_L/2$. In this frame, both optical and acoustic modes are precessing with frequency $\omega_L/2$ in opposite directions and thus have the same energy $\tilde E$ in Eq.~(\ref{MagnonSpectrumextendedRef}).

The two-component real field $(\theta_x,\theta_y)$  can be rewritten in terms of scalar complex field $\Phi$
\begin{equation}
\Phi =\left( \frac{\chi}{2} \right)^{1/2}
(\theta_x + i \theta_y)\,.
 \label{Phi}
\end{equation}
The Lagrangian (\ref{Lagrangian_theta2}) becomes the Lagrangian for a scalar field interacting with a $U(1)$ gauge field, whose time component is $A_0=\omega_L/2$:
\begin{align}
L= &- (\partial_t \Phi + i A_0 \Phi)(\partial_t \Phi^* - i A_0 \Phi^*) \nonumber \\
&+E_0^2  |\Phi|^2 + c^2 |\nabla \Phi|^2 + F_{\rm so}(\Phi,\Phi^*)
\,.
 \label{Lagrangian_Phi}
\end{align}
In a constant magnetic field, the $U(1)$ gauge is removed by introducing the time dependent phase rotation, $\tilde \Phi(t)=\Phi(t) e^{iMt}$, and one obtains the Klein-Gordon Lagrangian for the complex relativistic scalar field used for the description of $Q$-balls in high energy physics:
\begin{equation}
L=- |\partial_t \tilde\Phi |^2 +E_0^2  |\tilde\Phi|^2 + c^2 |\nabla \tilde\Phi|^2 + F_{\rm so}(\tilde\Phi,\tilde\Phi^*)
\,.
 \label{Lagrangian_Phi2}
\end{equation}
In transverse NMR, where transverse components of spins precess with the Larmor frequency $\omega_L$, the field $\tilde \Phi$ has the form:
\begin{equation}
\tilde \Phi(t)=\Phi(t) e^{iE_0t}=\Phi_0 e^{-i\omega_L t} e^{iE_0t}=\Phi_0e^{-iE_0t}
\,.
 \label{tilde_Phi_precessing}
\end{equation}

The energy spectrum of excitations of scalar field $\Phi$ in the absence of spin-orbit interaction is
\begin{equation}
\omega _\pm(k)=\pm \sqrt{E_0^2+k^2c^2} -  A_0\,,
 \label{ScalaSpectrum}
\end{equation}
which corresponds to the spin wave spectrum in Eq.~(\ref{MagnonSpectrumextendedM}).
The branch with minus sign gives the spectrum of optical magnons:  $E(S_z=-1,k)=|\omega _-(k)|$.


The Lagrangian (\ref{Lagrangian_Phi}) for the complex field in the absence of spin-orbit interaction has a conserved quantity -- the $U(1)$ charge $Q$:
\begin{equation}
Q= i\int d^3x \left(\tilde\Phi^* \partial_t \tilde\Phi  - \tilde\Phi \partial_t \tilde\Phi^* \right)
\,.
 \label{Q}
\end{equation}
In the precessing state (\ref{tilde_Phi_precessing}) one obtains:
\begin{equation}
Q= 2M\int dV ~|\Phi_0|^2=  \frac{\chi \omega_L}{2} \int dV ~\mbox{\boldmath$\theta$}_\perp^2
\approx   \int dV~ (S-S_z) 
\,,
 \label{Q_precession}
\end{equation}
where we used $S_z=\sqrt{S^2 - S_\perp^2}$ and $S_\perp^2=S^2 \mbox{\boldmath$\theta$}_\perp^2$.
Each magnon reduces the total spin by $\hbar$, and therefore the charge of the complex field coincides with the magnon number $N_{\rm M}$:
\begin{equation}
Q=N_{\rm M} 
\,, 
 \label{Q-magnons_number}
\end{equation}

\subsection*{From Klein-Gordon to Schr\"odinger equation}
We have shown that the dynamics of magnetisation obeys a relativistic Klein-Gordon equation, where the "speed of light" corresponds to the velocity of spin waves. The corresponding global $U(1)$ symmetry is the $SO_S(2)$ symmetry under rotation of spin system about the axis of applied constant magnetic field (axis $z$).
 The global $U(1)$ charge $Q$ comes from projection of spin along the direction of magnetic field, $Q=( S- S_z)/\hbar=N_{\rm M}$. This is a quasi-conserved quantity in $^3$He-B as magnons are long-lived quasiparticles.

The density of trapped magnons is relatively small and the direct interaction between them is negligible compared with the interaction with the flexible orbital texture. Consider the non-relativistic limit $kc \ll mc^2$ of the Klein-Gordon equation realised in the experiment. The wave vector $k$ is inverse of the characteristic length scale of the trapping potential $U({\bf r})$. As in this limit one has
\begin{equation}
\frac{(\omega -A_0)^2-E_0^2}{2E_0}\approx \omega -\omega_L 
\,,
 \label{mu}
\end{equation}
the Klein-Gordon equation for $\tilde\Phi$ transforms to Schr\"odinger equation for $\psi$:
 \begin{equation}
-i \hbar\frac{\partial \psi}{\partial t}= - \frac{\hbar^2}{2m}\Delta \psi + U\psi ~~,~~
U({\bf r})=\hbar\omega_L({\bf r})  + U_{\rm text}({\bf r})\,.
\label{Schrodinger}
\end{equation}
The Klein-Gordon wave function $\tilde \Phi$ is connected to the Schr\"odinger wave function $\psi$ for magnons
\begin{equation}
\tilde \Phi(t)=\frac{\psi}{\sqrt{2E_0}}  e^{iE_0t- i\omega t} ~~,~~\int dV ~|\psi|^2=N_{\rm M}
\,,
 \label{tilde_Phi_Schr}
\end{equation}
which satisfies $|\psi|^2=\omega_L|\Phi|^2$, and is normalised to the number of magnons: 
\begin{equation}
Q= i\int dV \left(\tilde\Phi^* \partial_t \tilde\Phi  - \tilde\Phi \partial_t \tilde\Phi^* \right)
=\int dV ~|\psi|^2=N_{\rm M} 
\,.
 \label{Normalization}
\end{equation}
The Schr\"odinger wave function  can be expressed in terms of observables, phase $\alpha_{\rm M}$ and tipping angle $\beta_{\rm M}$ of the precessing magnetisation:
 \begin{equation}
\psi = \sqrt{\frac{2\chi H}{\gamma \hbar}} \sin(\beta_{\rm M}/2)\ \exp(i\omega t + i\alpha_{\rm M}) \,.
\label{psi_vs_precession_full}
\end{equation}

The non-relativistic limit in Eq.~(\ref{MagnonSpectrum3}) of spectrum in Eq.~(\ref{MagnonSpectrumextendedM}) is obtained solving Eq.~(\ref{Schrodinger}) for a free particle, assuming the frequency of precession is close to the Larmor frequency, $|\omega-\omega_L| \ll \omega_L$.

The potential $U({\bf r})$ for magnons has two contributions: the spatial dependence of the local Larmor frequency $\omega_L({\bf r})=\gamma H({\bf r})$, and that of the spin-orbit interaction $F_{\rm so}({\bf r})$ in Eq.~(\ref{Lagrangian_Phi}) averaged over spin precession. The latter can be expressed in terms of the field of unit vector $\hat{\bf l}({\bf r})$, defined as the direction of the orbital angular momentum of Cooper pairs in $^3$He-B. The field of $\hat{\bf l}({\bf r})$-vector is time-independent in the spin precessing state:
\begin{equation} \label{TexturalPotential}
 F_{\rm so}=U_{\rm text} |\psi|^2,
\end{equation}
where
\begin{align} \label{TexturalPotential2}
& U_{\rm text}({\bf r})=\hbar \frac{2\Omega_B^2}{5\omega_L}\left(1-l_z({\bf r})\right) \nonumber \\ 
&= \hbar \frac{4\Omega_B^2}{5\omega_L}\sin^2(\beta_L({\bf r})/2)\equiv   \frac{1}{\omega_L} \zeta^2({\bf r})\,.
\end{align}
Here $\Omega_B\ll \omega_L$ is the so-called Leggett frequency which characterises the magnitude of the spin-orbit interaction \cite{VW}, and $\beta_L$ is the polar angle of the $\hat{\bf l}$-vector. The texture of the polar angle $\beta_L({\bf r})$ plays the role of the neutral scalar field $\zeta({\bf r})$ interacting with the complex field $\Phi({\bf r})$ (or $\psi({\bf r})$). The texture $\zeta({\bf r})$ is obtained by minimisation of the textural energies \cite{thuneberg_texture, kopu_texture} with addition of the contribution which comes from the complex field of magnons \cite{magnon_trap_mod}:
\begin{equation}
F_{\rm so}=    \frac{1}{\omega_L} |\psi({\bf r})|^2 \zeta^2({\bf r}) \equiv    |\Phi({\bf r})|^2 \zeta^2({\bf r})\,.
 \label{InteractionIntroduction}
\end{equation}

\section*{Appendix B: Magnon $Q$-ball and MIT hadron bag}

 \begin{figure}[tb]
 \centerline{\includegraphics[width=1.0\linewidth]{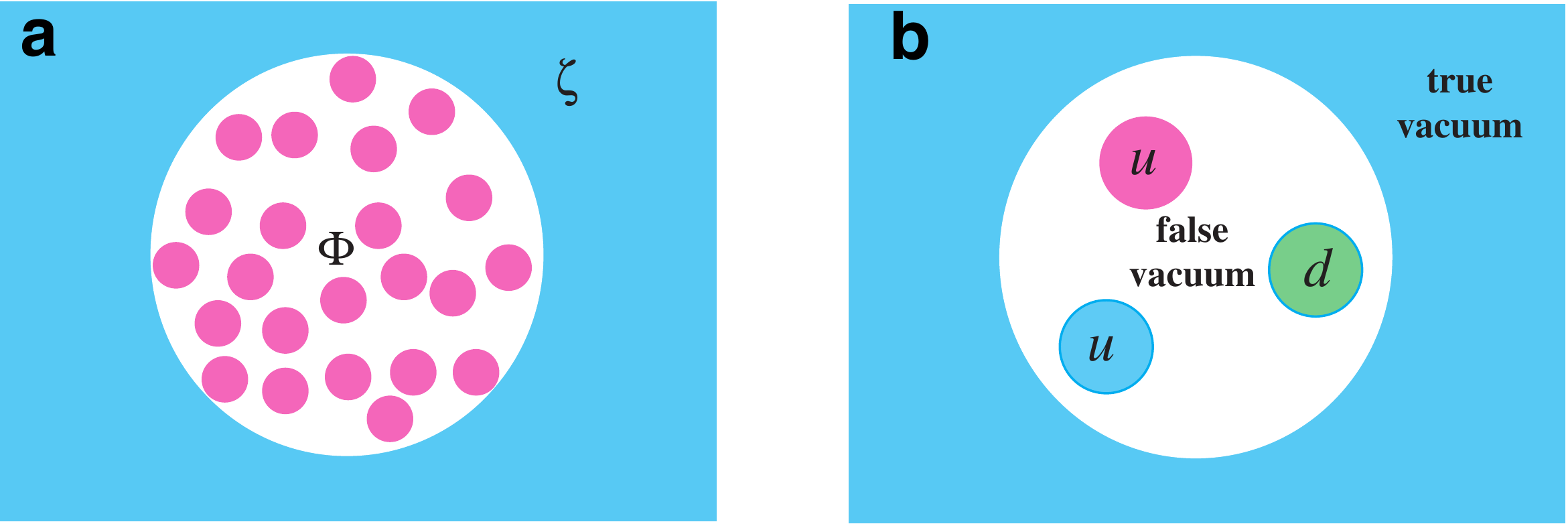}}
   \caption{\label{Fig:MIT} $Q$-ball in $^3$He-B as MIT hadron bag.  ({\bf a}) The BEC of magnons (red circles) in the limit of large $N_{\rm M} $ in a box cavity, which is void of neutral field and filled with the charge field $\Phi({\bf r})$. ({\bf b}) Simplified MIT bag model of hadrons, where the quarks forming a hadron are confined within a blob of false vacuum void of the actual vacuum of the QCD field.
 }
 \end{figure}
 
The magnon $Q$-ball in $^3$He-B is formed owing to interaction between the magnon condensate described by the charged field $\Phi$ with conserved charge $Q=N_{\rm M} $, and the orbital field $\zeta({\bf r})$, which is an analogue of the neutral field \cite{volovik_bunkov_qball,friedberg_qball}. The field $\zeta({\bf r})$ forms a potential well in which the charge $Q=N_{\rm M} $ is condensed. In the process of self-localization the charged field $\Phi({\bf r})$ modifies locally the neutral field $\zeta({\bf r})$ via the spin-orbit interaction (Eq. (\ref{InteractionIntroduction})). That interaction is repulsive and, if the magnetic part of the trapping potential $U_H$ is neglected, in the limit of large $N_{\rm M} $ a cavity is formed, which is void of neutral field $\zeta({\bf r})$ and filled with the charge field $\Phi({\bf r})$. That is, the flexible textural trap $U_\mathrm{text}$ transforms to a box with walls impenetrable for magnons \cite{magnon_trap_mod}. The pressure from the field $\zeta$ is 
compensated by the zero point pressure of the free magnons. This is an analogue of the MIT bag model of hadrons, where the quarks forming a hadron are confined only within the QCD vacuum field, and the quarks can can freely move in the false vacuum void of the QCD field \cite{rota2012}. The confined quarks form a blob of false vacuum, where the external pressure form the QCD vacuum is compensated by the zero point pressure of the confined quarks.

\section*{Appendix C: Symmetry breaking}

\begin{figure}[th!]
 \centerline{\includegraphics[width=\linewidth]{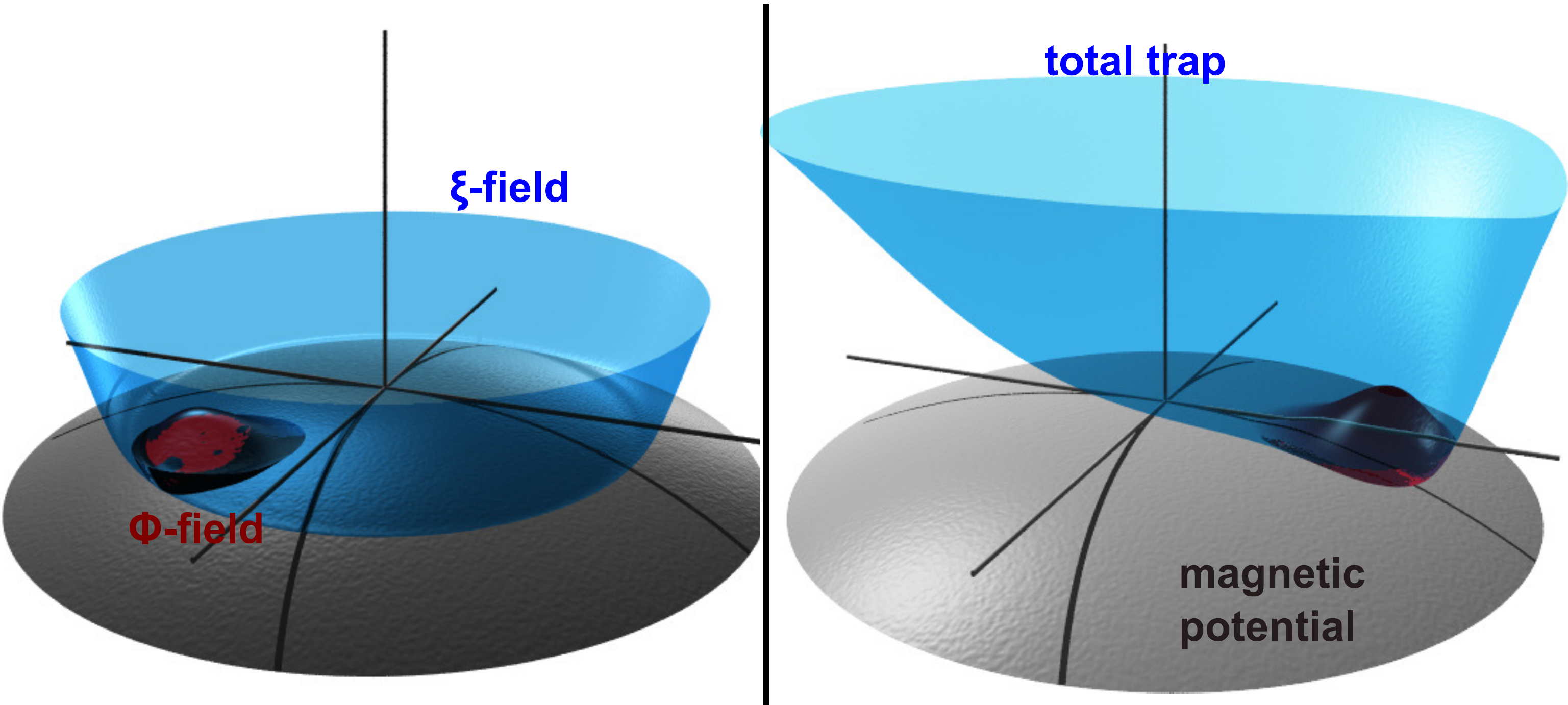}}
 \caption{\label{Fig:Hat}
   Illustration of a conventional ({\it left}) and an unconventional ({\it right}) spontaneous breaking of global continuous symmetry: In conventional situation, the potential acquires the form of a Mexican hat, but remains axisymmetric. The trapped particle(s) becomes localised in one of the degenerate points in the valley of the potential, thus breaking the $U(1)$-symmetry. For magnon $Q$-ball the situation is different: The Mexican hat potential itself is unstable towards breaking of axial symmetry due to the self-trapping effect. The $\zeta$-field conforms the movements of the magnon $Q$-ball soliton ($\Phi$-field).
 }
 \end{figure}

Let us compare the $Q$-ball fromation with conventional symmetry breaking, such as the symmetry breaking which triggers the Higgs mechanism in the Standard model \cite{Nakamura2010}. In the conventional case the potential acquires the shape of a Mexican hat, but remains axisymmetric as in Fig. \ref{Fig:Hat} {\it left}. In our case the potential $U_{\rm text}({\bf r})$ does not have the Mexican hat shape, Fig. \ref{Fig:Hat} {\it right}:  The potential shape depends on density of bosons localised in it. Therefore the axisymmetric Mexican hat potential itself is unstable towards symmetry breaking in the azimuthal coordinate. Although the generalised Hamiltonian for the combined $\Phi$ and $\hat{\bf l}$ fields remains symmetric (degenerate), in the $Q$-ball picture the potential  $U_{\rm text}({\bf r})$ is localised along the bosons. This is a unique experimental example of spontaneous breaking of the rotational SO(2) symmetry on top of formation of the axisymmetric Mexican hat potential. 

%

\end{document}